\documentstyle[preprint,aps,floats,psfig]{revtex}
\begin{document}
\title  {Charge current in ferromagnet - triplet superconductor junctions}

\author{ N. Stefanakis}
\address{ Department of Physics, University of Crete,
	P.O. Box 2208, GR-71003, Heraklion, Crete, Greece}

\date{\today}
\maketitle
\begin{abstract}

We calculate the tunneling conductance spectra of a 
ferromagnetic metal / insulator / triplet superconductor 
from the reflection amplitudes using the 
Blonder-Tinkham-Klapwijk (BTK) formula. 
For the triplet superconductor, we assume one special 
$p$-wave order parameter, having line nodes, and two 
two dimensional $f$-wave order parameters with line 
nodes, breaking the time reversal symmetry. 
Also we examine nodeless pairing potentials.
The evolution of the spectra with the exchange potential 
depends solely on the topology of the gap.
The weak Andreev reflection within the ferromagnet results
in the suppression of the tunneling conductance and 
eliminates the resonances due to the anisotropy of the 
pairing potential.
The tunneling spectra splits asymmetrically with respect 
to $E=0$ under the influence of an external magnetic field.
The results can be used to 
distinguish between the possible candidate pairing states of 
the superconductor Sr$_2$RuO$_4$.

\end{abstract}


\section{Introduction}
The recent discovery of superconductivity in Sr$_2$RuO$_4$ has 
attracted much theoretical and experimental interest \cite{maeno}.
The time reversal symmetry is broken 
for the superconductor Sr$_2$RuO$_4$, and the magnetic field
is spontaneously induced as shown by $\mu$SR experiment. \cite{luke}.
The Knight-shift shows no change when passing 
through the superconducting state and is a clear 
indication for spin triplet pairing state, with a 
${\bbox d}$-vector aligned to the $z$-axis.
\cite{ishida}.
In addition the band structure calculations 
\cite{mazin} and de Haas van
Alphen measurements \cite{mackenzie} show little dispersion along 
$k_z$ which is consistent with a two dimensional basis
function on a cylindrical Fermi surface.
Furthermore, the presence of a large 
residual density of states of quasiparticles 
inside the superconducting gap is evident from 
the linear temperature dependence of the nuclear spin 
lattice relaxation rate $1/T_1$ of $^{101}$Ru bellow $0.4K$. 
\cite{ishida1}
Also specific heat measurements support the scenario of 
line nodes within the 
gap as in the high $T_c$ cuprate superconductors \cite{nishizaki}.

In the tunneling experiments between normal metals and superconductors
Andreev reflection process take place \cite{blonder,andreev}.
In the Andreev reflection process an electron incident, 
in the barrier with an energy bellow the gap 
can not drain off into the superconductor. It is instead reflected 
as a hole and a Cooper pair is transfered into the superconductor.
In anisotropic high $T_c$ superconductors 
due to the sign change of the pair potential that 
the transmitted quasiparticles feel,
zero energy states are formed, which are detected as peaks in the
conductance spectra at $E=0$.
\cite{tanaka1}
Also in the presence of an imaginary $s$-wave component, which 
breaks the time reversal symmetry, the zero energy peak is shifted to the 
amplitude of the subdominant component.
\cite{covington,stefan1}  

The properties of the Andreev reflection are modified in the 
presence of an exchange field as in ferromagnet / insulator / superconductor
junctions, since the retro-reflection of the Andreev reflection is 
broken in the ferromagnet. This phenomenon has been 
clarified both in $s$-wave and $d$-wave junctions and 
interesting aspects of the Andreev reflection have been revealed.
\cite{dimoulas,jong,kashiwaya,zhu}
Also the properties of ferromagnet / insulator / triplet superconductor 
junctions have been studied where two types of pairing potentials 
are assumed for the triplet superconductor, i.e. the unitary 
and the non unitary with $E_u$ symmetry. \cite{yoshida}
In the unitary case the 
conductance within the gap is reduced with the exchange interaction 
while in the non unitary it is not much 
influenced since the latter pairing state conserves spin.

In order to identify the pairing state of Sr$_2$RuO$_4$
the Bogoliubov-de Gennes (BdG) equations have been used 
to calculate the quasiparticle bound state wave function around    
non-magnetic impurities in unconventional superconductors. \cite{maki}
The characteristic patterns were distinguished 
for two proposed order parameters, 
(i) $E_u$, and 
(ii) $B_{1g}\times E_u$. 

In this paper we will use the
BdG equations to calculate the 
tunneling conductance of ferromagnet  /
triplet superconductor contacts, with a barrier of arbitrary strength 
between them, in terms of the 
probability amplitudes of Andreev and normal 
reflection.
For the triplet superconductor we shall assume 
three possible pairing states of two 
dimensional order parameters,
having line nodes within the RuO$_2$ plane, which 
break the time reversal symmetry.
The first two are the $2D$ $f$-wave states 
proposed by Hasegawa et al,
\cite{hasegawa} having $B_{1g}\times E_u$ and $B_{2g}\times E_u$ symmetry
respectively.
The other one is called nodal $p$-wave state and has been 
proposed by Dahm et al \cite{dahm}, where 
the pairing potential has the form 
${\bbox d}({\bbox k})=\Delta_0 \hat{\bbox z} (\sin(k_xa)+i\sin(k_ya))$, with 
$k_xa=\pi \cos\theta$, $k_ya=\pi \sin\theta$. This pairing 
symmetry has nodes as in the $B_{2g}\times E_u$ case.
Also we will consider two nodeless pairing states. One is 
the isotropic $p$-wave state and the other is the 
nodeless $p$-wave state initially proposed by 
K. Miyake and O. Narikiyo \cite{miyake}, both breaking 
the time reversal symmetry.
Generally the tunneling conductance is suppressed 
with the increase of the exchange interaction and the peaks 
are removed. This is due to the suppression of the Andreev 
reflection in the ferromagnet. For the nodal pairing states the linear 
dependence of the tunneling conductance with $E$ is not 
much influenced. For the nodeless cases, 
the normalized conductance develops 
a constant value within the gap, which is suppressed 
as the exchange field gets larger.
When the ferromagnet is a normal  metal,
the magnetic field splits the tunneling spectrum 
symmetrically around $E=0$.
The exchange field eliminates the negative branch of the 
tunneling spectra in the half metallic ferrogmagnetic 
limit.

\section{Theory of tunneling effect}

For spin-triplet superconductors the wave functions describing 
the quasiparticles $\hat{\Psi}({\bbox r})$ are 
four-spinors in Nambu (particle-hole $\otimes$ spin) space.
Their particle and hole components are determined by the 
solutions of the BdG equations
\cite{ueda,honerkamp}
\begin{equation}
{
E\hat{\Psi}({\bbox r})=\int d{\bbox r'}
\hat{H}({\bbox r},{\bbox r'})
\hat{\Psi}({\bbox r'})
},~~~\label{bgdfour}
\end{equation}
where,
\begin{equation}
\hat{\Psi}({\bbox r})=
\left(
\begin{array}{l}
 u_{\uparrow}({\bbox r}) \\
 u_{\downarrow}({\bbox r}) \\
 v_{\uparrow}({\bbox r}) \\
 v_{\downarrow}({\bbox r}) 
\end{array}
\right)
,~~~\label{psifour}
\end{equation}

\begin{equation}
\hat{H}({\bbox r},{\bbox r'}) = 
  \left(
    \begin{array}{ll}
     \hat{H_e} &   
     \hat{\Delta} \\
     \hat{\Delta}^{\ast} &
     -\hat{H_e} 
    \end{array}
  \right)
.~~~\label{deltafour}
\end{equation}
$\hat{\Delta}$ is the $2 \times 2$ triplet pairing matrix with
elements of the form $\Delta_{s\overline{s}}({\bbox r},{\bbox r'})$, 
and the spin index $s$=$\uparrow$, or $s=\downarrow$.
$\hat{H_e}=H_e({\bbox r'}) 
\delta({\bbox r}-{\bbox r'}) \hat{\sigma_0}$, where $\hat{\sigma_0}$ 
is the $2 \times 2$ unit matrix,
and ${\cal H}_e(
{\bbox r})$ is the single-particle Hamiltonian which is given by ${\cal H}_e({\bbox r})=
-\hbar^2\bigtriangledown_{\bbox r}^2/2m_e+V({\bbox r})-E_F$, $E$ is the energy 
measured from the Fermi energy $E_F$.
For the pairing states we examine in Sec. III, the spin-up and spin-down 
components decouple. 
We will consider only triplet pairing states 
where 
$\Delta_{\uparrow \uparrow}({\bbox r},{\bbox r'}) =
\Delta_{\downarrow \downarrow}({\bbox r},{\bbox r'}) =0$, while
$\Delta_{\uparrow \downarrow}({\bbox r},{\bbox r'}) =
\Delta_{\downarrow \uparrow}({\bbox r},{\bbox r'})$. 
In that case the cooper pairs have zero spin projection.
The spin dependent BdG equations are decoupled into two independent sets of 
(two component) equations, one for the spin up electron, 
spin down hole quasiparticle $(u_{\uparrow}({\bbox r}),v_{\downarrow}({\bbox r}))$, 
and the other for $(u_{\downarrow}({\bbox r}),v_{\uparrow}({\bbox r}))$. 
The corresponding BdG equations for spin index 
$s(\overline{s})=\uparrow(\downarrow)$ or 
$s(\overline{s})=\downarrow(\uparrow)$, read
\cite{bruder}

\begin{equation}
{
\begin{array}{ll}
Eu_s({\bbox r}) & = ({\cal H}_e({\bbox r})-\rho U({\bbox r})) u_s({\bbox r})+
\int d{\bbox r'}\Delta_{s\overline{s}}({\bbox s},{\bbox x})v_{\overline{s}}
({\bbox r'}) \\
Ev_{\overline{s}}({\bbox r}) & = -({\cal H}_e^{\ast}({\bbox r})+\rho U({\bbox r}))
v_{\overline{s}}({\bbox r})+
\int d{\bbox r'}\Delta_{\overline{s}s}^{\ast}({\bbox s},{\bbox x}) u_{s}({\bbox r'})
\end{array},~~~\label{bdg}
}
\end{equation}
where $U({\bbox r})$ is the exchange potential, $\rho$ is $1(-1)$ 
for up(down) spins.
$\Delta_{s\overline{s}}({\bbox s},{\bbox x})$ 
is the matrix element of the pair potential, after a transformation 
from the position coordinates ${\bbox r},{\bbox r'}$ to the center of mass 
coordinate ${\bbox x}=({\bbox r}+{\bbox r'})/2$ 
and the relative vector ${\bbox s}={\bbox r}-{\bbox r'}$. 
After Fourier transformation the pair potential depends on the 
related wave vector ${\bbox k}$ and ${\bbox x}$. 
In the weak coupling limit ${\bbox k}$ is 
fixed on the Fermi surface ($|{\bbox k}|=k_F$), and only its direction $\theta$ is 
variable.
After applying the quasi-classical approximation, i.e.
\begin{equation}
\left(
\begin{array}{ll}
 \overline{u}_s({\bbox r}) \\
 \overline{v}_{\overline{s}}({\bbox r}) 
\end{array}
\right)
=e^{-i{\bbox k} \cdot {\bbox r}}
\left(
\begin{array}{ll}
  u_s({\bbox r}) \\
  v_{\overline{s}}({\bbox r}) 
\end{array}
\right)
,~~~\label{out}
\end{equation}
so that the 
fast oscillating part, of the wave function is divided out, the BdG equations
are reduced to the Andreev equations \cite{andreev}
\begin{equation}
{
\begin{array}{ll}
E\overline{u}_s({\bbox r}) & = -i\hbar^2/m{\bbox k} \cdot {\bbox \bigtriangledown} 
\overline{u}_s({\bbox r})+
\Delta_{s\overline{s}}(\theta,{\bbox r})\overline{v}
_{\overline{s}}({\bbox r}) \\
E\overline{v}_{\overline{s}}({\bbox r}) & = i\hbar^2/m{\bbox k} \cdot {\bbox \bigtriangledown}
\overline{v}_{\overline{s}}({\bbox r})+
\Delta_{\overline{s}s}^{\ast}(\theta,{\bbox r}) \overline{u}_{s}({\bbox r})
\end{array},~~~\label{ndrv}
}
\end{equation}
where the quantities $\overline{u}_s({\bbox r})$ and $\overline{v}_{\overline{s}}({\bbox r})$
are electron-like and hole-like quasiparticles with spin index 
$s$, and $\overline{s}$ respectively. 

\begin{figure}
  \centerline{\psfig{figure=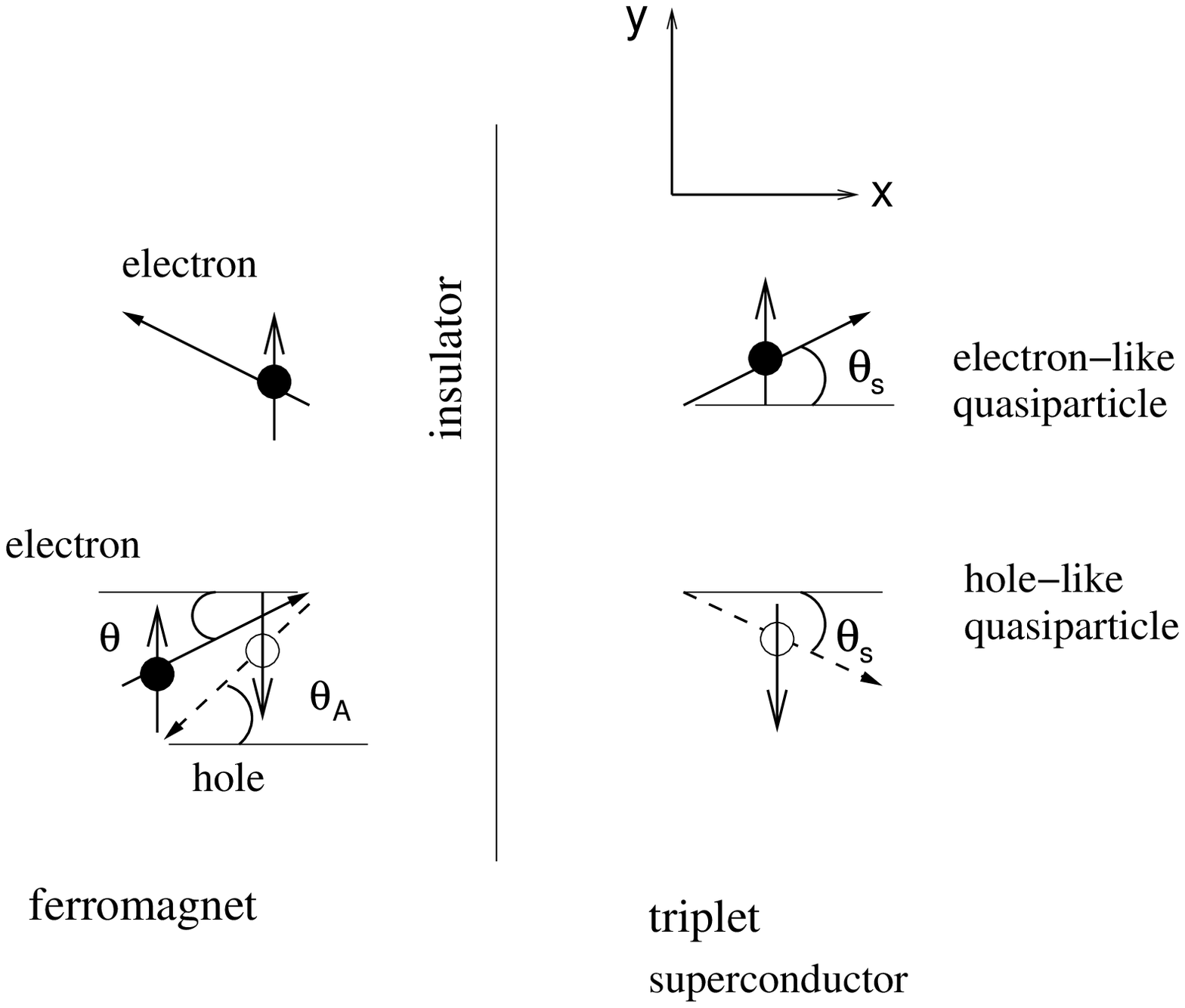,width=8.5cm,angle=0}}
  \caption{
The geometry of the ferromagnet / insulator / triplet superconductor 
interface. The pairing state is unitary with zero spin projection. 
The vertical line along the $y$-axis represents the 
insulator. 
The arrows illustrate the transmition and reflection processes at the 
interface.
$\theta$ is the angle of the incident electron and the normal,
$\theta_A$ is the angle of the reflected hole and the normal, and 
$\theta_s$ is the angle of the transmitted quasiparticle and the normal.
Note that $\theta$ is not equal to $\theta_A$ since the retroreflection of the 
Andreev process is lost.
}
  \label{fig1.fig}
\end{figure}

We consider the ferromagnet / insulator / superconductor 
junction shown in Fig. \ref{fig1.fig}.
The geometry of the problem has the following limitations.
The particles move in the $xy$-plane
and the boundary between the ferromagnet 
($x<0$) and 
superconductor ($x>0$) is the $yz$-plane at $x=0$. 
The insulator is modeled by a delta function, located at $x=0$, of the 
form $V\delta(x)$. The temperature is fixed to $0$ K.
We take both the pair potential and the exchange energy 
as a step function i.e.
$\Delta_{s\overline{s}}(\theta,{\bbox r})=\Theta(x)\Delta_{s\overline{s}}(\theta)$, 
$U{\bbox r})=\Theta(-x)U$.
For the geometry shown in Fig. \ref{fig1.fig}, Eqs. \ref{ndrv} 
take the form 
\begin{equation}
{
\begin{array}{ll}
E\overline{u}_s(x) & = -i\hbar^2/m k_{Fx} \frac{d}{dx}
\overline{u}_s(x)+
\Delta_{s\overline{s}}(\theta)\overline{v}
_{\overline{s}}(x) \\
E\overline{v}_{\overline{s}}(x) & = i\hbar^2/m k_{Fx} \frac{d}{dx}
\overline{v}_{\overline{s}}(x)+
\Delta_{\overline{s}s}^{\ast}(\theta) \overline{u}_{s}(x)
\end{array}.~~~\label{ndrv1d}
}
\end{equation}

When a beam of electrons is incident from the ferromagnet
to the insulator, with an angle $\theta$, the general solution 
of Eqs. (\ref{ndrv1d}), is the two 
component wave function $\Psi_I=(u_{\uparrow[\downarrow]},
v_{\downarrow[\uparrow]})$ which 
for $x<0$ is written as
\begin{equation}
\Psi_I=
\left(
\begin{array}{ll}
  1 \\
  0 
\end{array}
\right)
e^{iq_{\uparrow [\downarrow]}x\cos\theta}+a_{\uparrow [\downarrow]}
\left(
\begin{array}{ll}
  0 \\
  1 
\end{array}
\right)
e^{iq_{\downarrow [\uparrow]}x\cos\theta_A}+b_{\uparrow [\downarrow]}
\left(
\begin{array}{ll}
  1 \\
  0 
\end{array}
\right)
e^{-iq_{\uparrow [\downarrow]}x\cos\theta},
~~~\label{x_}
\end{equation}
where $a_{\uparrow [\downarrow]},b_{\uparrow [\downarrow]}$, 
are the amplitudes for Andreev and normal reflection
for spin up(down) quasiparticles, and 
$q_{\uparrow [\downarrow]}=\sqrt{\frac{2 m}{\hbar^2}(E_F\pm U)}$
is the wave vector of quasiparticles in the ferromagnet for 
up (down) spin.
The wave vector of the electron-like, hole-like quasiparticles 
is approximated by $k_s=\sqrt{\frac{2 m E_F}{\hbar^2}}$.
Since the translational symmetry holds in the $y$-axis 
direction, the momenta parallel to the interface is conserved, 
i.e. $q_{\uparrow}\sin\theta=q_{\downarrow}\sin\theta_A=k_s\sin\theta_s$.
Note that $\theta$ is different than $\theta_A$ since the 
retroreflection of the Andreev reflection is broken.
Using the matching conditions of the wave function at $x=0$,
$\Psi_I(0)=\Psi_{II}(0)$ and 
$\Psi_{II}'(0)-\Psi_{I}'(0)=(2mV/\hbar^2)\Psi_I(0)$, 
the Andreev and normal reflection amplitudes
$a_{\uparrow [\downarrow]},b_{\uparrow [\downarrow]}$
for the spin up(down) quasiparticles are obtained

\begin{equation}
a_{\uparrow [\downarrow]}=\frac{4n_{+}\lambda_1}
     {(-1-\lambda_1-iz_{\uparrow [\downarrow]})
      (-1-\lambda_2+iz_{\uparrow [\downarrow]})+
     (1-\lambda_1-iz_{\uparrow [\downarrow]})
     (-1+\lambda_2-iz_{\uparrow [\downarrow]})n_{+}n_{-}
      \phi_{-}\phi_{+}^{\ast}}
,~~~\label{a}
\end{equation}

\begin{equation}
b_{\uparrow [\downarrow]}=\frac
     {(-1-\lambda_2+iz_{\uparrow [\downarrow]})
      (1-\lambda_1+iz_{\uparrow [\downarrow]})+
     (-1+\lambda_2-iz_{\uparrow [\downarrow]})
     (-1-\lambda_1+iz_{\uparrow [\downarrow]})n_{+}n_{-}
      \phi_{-}\phi_{+}^{\ast}}
     {(-1-\lambda_1-iz_{\uparrow [\downarrow]})
      (-1-\lambda_2+iz_{\uparrow [\downarrow]})+
     (1-\lambda_1-iz_{\uparrow [\downarrow]})
     (-1+\lambda_2-iz_{\uparrow [\downarrow]})n_{+}n_{-}
      \phi_{-}\phi_{+}^{\ast}}
,~~~\label{b}
\end{equation}
where $z_0=\frac{m V}{\hbar^2 k_s}$, 
$z_{\uparrow [\downarrow]}=\frac{2 z_0}{\cos\theta_s}$, 
$\lambda_1=\frac{\cos\theta}{\cos\theta_s}
\frac{q_{\uparrow [\downarrow]}}{k_s}$,
$\lambda_2=\frac{\cos\theta_A}{\cos\theta_s}
\frac{q_{\downarrow [\uparrow]}}{k_s}$.
The BCS coherence factors are given by 
\begin{equation}
u_{\pm}^2=[1+
      \sqrt{E^2-|\Delta_{\pm}(\theta)|^2}/E]/2,
\end{equation}

\begin{equation}
v_{\pm}^2=[1-
      \sqrt{E^2-|\Delta_{\pm}(\theta)|^2}/E]/2,
\end{equation}
and $n_{\pm}=v_{\pm}/u_{\pm}$.
The internal phase coming from the energy gap is given by
$\phi_{\pm} =[
\Delta_{\pm}(\theta)/|\Delta_{\pm}(\theta)|]$,
where $\Delta_{+}(\theta)=\Delta(\theta)$
($\Delta_{\_}(\theta)=\Delta(\pi- \theta)$), is the 
pair potential experienced by the transmitted electron-like 
(hole-like) quasiparticle.
$\Delta(\theta)=\Delta_{\uparrow \downarrow}(\theta)$
$=\Delta_{\downarrow \uparrow}(\theta)$, since the cooper 
pairs have zero spin projection i.e. ${\bbox d} \parallel \hat {\bbox z}$.

When $\theta > \sin^{-1}(\frac{k_s}{q_{\uparrow}})\equiv \theta_{c1}$ 
total reflection occurs and the spin and charge current vanishes.
When $\theta_{c1} > \theta > \sin^{-1}(\frac{q_{\uparrow}}{q_{\downarrow}})$ 
$\equiv \theta_{c2}$, although the transmitted quasiparticles 
in the superconductor do propagate, the Andreev reflected quasiparticles 
do not propagate. This process is called virtual Andreev reflection 
(VAR process). In this case the spin and charge current do not vanish
since a finite amplitude of the Andreev reflection still exists.
\cite{kashiwaya}

According to the BTK formula the conductance for the charge current
of the junction, 
$\overline{\sigma}_{q_{\uparrow [\downarrow]} }(E,\theta)$, 
for up(down) spin quasiparticles, 
is expressed in terms of the 
probability amplitudes
$a_{\uparrow [\downarrow]},b_{\uparrow [\downarrow]}$ as
\cite{blonder,kashiwaya}
\begin{equation}
\overline{\sigma}_{q_{\uparrow [\downarrow]}}(E,\theta) 
=Re\left [1+\frac{\lambda_2}{\lambda_1}|a_{\uparrow [\downarrow]}|^2
-|b_{\uparrow [\downarrow]}|^2\right ]
.~~~\label{ovs}
\end{equation}
The tunneling conductance, normalized by that in the normal 
state is given by 

\begin{equation}
\sigma_q(E)=
\sigma_{q_{\uparrow }}(E)+
\sigma_{q_{\downarrow }}(E)
,~~~\label{sqcharge}
\end{equation}
	
\begin{equation}
\sigma_{q_{\uparrow [\downarrow]}}(E)=
	\frac{1}{R_N}
\int_{-\pi/2}^{\pi/2}d\theta \cos \theta 
\overline{\sigma}_{q_{\uparrow [\downarrow]}}(E,\theta)
	P_{\uparrow [\downarrow]}q_{\uparrow [\downarrow]}
,~~~\label{sq}
\end{equation}
where
\begin{equation}
R_N=
\int_{-\pi/2}^{\pi/2}d\theta \cos \theta [ \sigma_{N_{\uparrow}}(\theta)
	P_{\uparrow}q_{\uparrow}+
\sigma_{N_{\downarrow}}(\theta)
	P_{\downarrow}q_{\downarrow}]
,~~~\label{RN}
\end{equation}

\begin{equation}
\sigma_{N_{\uparrow [\downarrow]}}(\theta)
      =\frac{4\lambda_1}{(1+\lambda_1)^2+z_{\uparrow [\downarrow]}^2}
,~~~\label{sN}
\end{equation}
where $P_{\uparrow [\downarrow]}=(E_F\pm U)/2E_F$ is the 
polarization for up(down) spin.
In the $z_0=0$ limit the interface is 
regarded as a weak link, showing metallic behavior 
while for large $z_0$
values the interface becomes insulating.

\section{Possible spin triplet pairing states}
For the spin triplet pairing state the Cooper pairs have 
spin $1$ degree of freedom. 
The gap function is a $2\times 2$ symmetric matrix which in the 
spin space can 
be written as 
\begin{equation}
\hat{\Delta}({\bbox k})= i\sigma_y 
({\bbox d}({\bbox k})\cdot \hat{\bbox \sigma} ),
\end{equation}
where $\hat{\bbox \sigma}$ denotes the Pauli matrices and 
${\bbox d}({\bbox k})$ is a vectorial function which is odd in
${\bbox k}$. The ${\bbox d}$ vector defines the axis along which 
the Cooper pairs have zero spin projection. In the following 
we will take 
${\bbox d} \parallel \hat{\bbox z}$. 
In that case
$\Delta_{\uparrow \uparrow} =
\Delta_{\downarrow \downarrow} =0$, while  
$\Delta_{\uparrow \downarrow} =
\Delta_{\downarrow \uparrow} =\Delta(\theta)$.
The energy spectrum of the quasiparticles consist of two 
branches which are identical for unitary pairing states,  
i.e. $\hat{\bbox \Delta}^{\dagger}({\bbox k})\hat{\bbox \Delta}({\bbox k})$ 
is proportional to the unit matrix, 
and distinct for non unitary states. 
The non-unitary states have been ruled out for Sr$_2$RuO$_4$
by the vary small residual
value of the specific heat at zero temperature \cite{nishizaki}.
In this paper we will examine 
only the case of unitary pairing states. 
As an example we consider the state
\begin{equation}
 \hat{\Delta} (\theta) = \Delta_0 
  \left(
    \begin{array}{ll}
     \Delta_{\uparrow \uparrow}(\theta)  &   
     \Delta_{\uparrow \downarrow}(\theta)    \\
      \Delta_{\downarrow \uparrow}(\theta)
       & \Delta_{\downarrow \downarrow}(\theta) 
    \end{array}
  \right)
.~~~\label{delta22}
\end{equation}

a) For the isotropic $p$-wave pairing state 
$\Delta_{\uparrow \downarrow}(\theta) =
\Delta_{\downarrow \uparrow}(\theta) =\Delta_0exp(i(\theta-\beta))$, 
and 
$\Delta_{\uparrow \uparrow}(\theta) =
\Delta_{\downarrow \downarrow}(\theta) = 0$,
$\beta$ denotes the angle between the normal to the interface
and the $x$-axis of the crystal.
This is an opposite spin pairing state, with a gap 
of constant modulus for both spin parts on the Fermi surface. 
In the following we will consider the cases where the matrix element 
$\Delta_{\uparrow \downarrow}(\theta)$ (expressed as $\Delta(\theta)$),
of Eq. \ref{delta22} has the following $\theta$ dependences.

b) In case of a $p$-wave superconductor, proposed by K. Miyake, and O. Narikiyo
\cite{miyake}
\begin{equation}
\Delta(\theta)=
\frac{\Delta_0}{s_M}[\sin(k_xa)+i\sin(k_ya)]
,~~~\label{MN}
\end{equation}
with $k_xa=R \pi \cos(\theta-\beta)$, and $k_ya=R \pi \sin(\theta-\beta)$,
$s_M=\sqrt{2}\sin\frac{\pi}{\sqrt{2}}=1.125$, and $R=0.9$.
This state does not has nodes.

We consider also 
three pairing symmetries for Sr$_2$RuO$_4$ with line nodes.
 
c) In the first $2D$ $f$-wave state $B_{1g}\times E_u$
\begin{equation}
\Delta(\theta)=\Delta_0
\cos2(\theta - \beta)[\cos(\theta - \beta) + i\sin(\theta - \beta)]
.~~~\label{cos}
\end{equation}
This state has nodes at the same points as in the $d_{x^2-y^2}$-wave
case.

d) For the second $2D$ $f$-wave state $B_{2g}\times E_u$
\begin{equation}
\Delta(\theta)=\Delta_0
\sin2(\theta - \beta)[\cos(\theta - \beta) + i\sin(\theta - \beta)]
.~~~\label{sin}
\end{equation}
This state has nodes at $0, \pi/2, \pi, 3\pi/2$, and has also 
been studied by Graf and Balatsky \cite{graf}.

e) In case of a nodal $p$-wave superconductor
\begin{equation}
\Delta(\theta)=
\frac{\Delta_0}{s_M}[\sin(k_xa)+i\sin(k_ya)]
,~~~\label{nodalp}
\end{equation}
with $k_xa=\pi \cos(\theta-\beta)$, and $k_ya=\pi \sin(\theta-\beta)$.
We use here the same normalization proposed by Dahm et al \cite{dahm}
$s_M=\sqrt{2}\sin\frac{\pi}{\sqrt{2}}=1.125$, where the 
Fermi wave vector is chosen as $k_Fa=\pi$, in order to have 
a node in $\Delta(\theta)$. 
This state has nodes as in the $B_{2g}\times E_u$ state.
The corresponding nodeless form was initially 
proposed by K. Miyake, and O. Narikiyo
\cite{miyake} and is considered as a separate case.

\section{Tunneling conductance characteristics} 

\begin{figure}
  \centerline{\psfig{figure=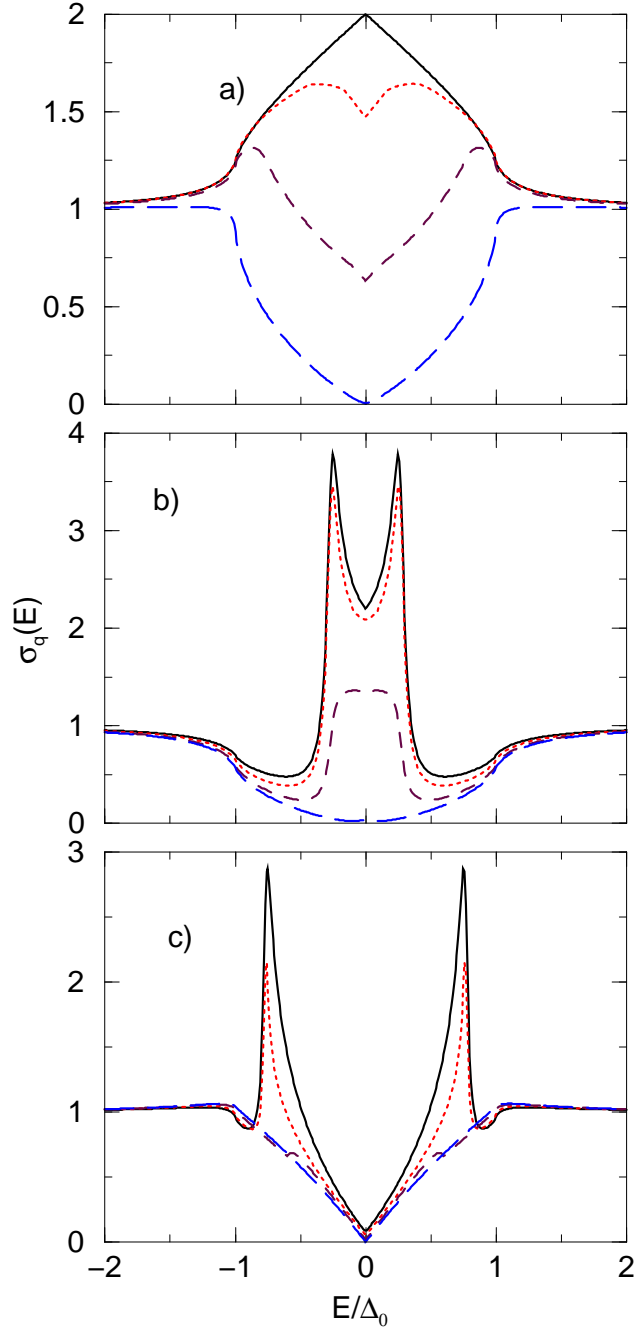,width=8.5cm,angle=0}}
  \caption{
Normalized tunneling conductance $\sigma_q(E)$ as a function of $E/\Delta_0$
for $x=0$ (solid line), $x=0.4$ (dotted line), $x=0.8$ (dashed line), 
and $x=0.999$ (long dashed line), 
for different orientations (a) Z=0, $\beta=0$, (b)$Z=2.5$,  $\beta=0$, 
(c) $Z=2.5$,  $\beta=\pi/4$. 
The pairing 
symmetry of the superconductor is 
$B_{1g}\times E_u$.
}
  \label{cos.fig}
\end{figure}

\begin{figure}
  \centerline{\psfig{figure=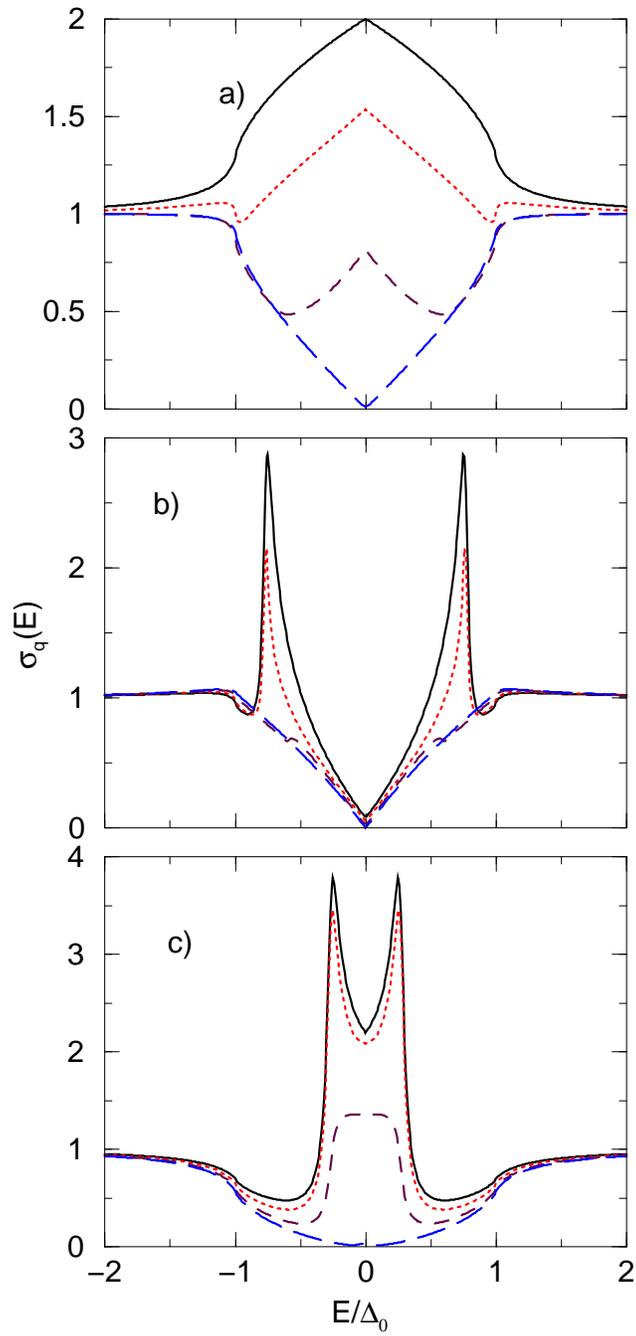,width=8.5cm,angle=0}}
  \caption{
The same as in Fig. 2. The pairing
symmetry of the superconductor is
$B_{2g}\times E_u$.
}
  \label{sin.fig}
\end{figure}

\begin{figure}
  \centerline{\psfig{figure=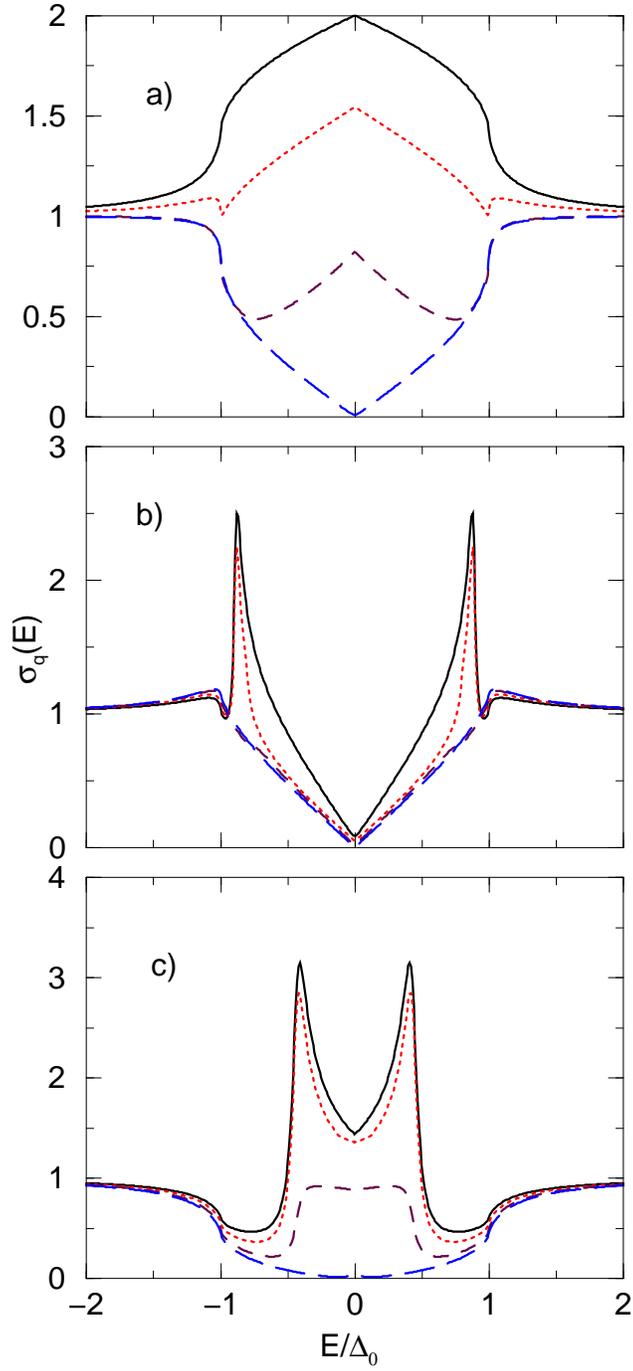,width=8.5cm,angle=0}}
  \caption{
The same as in Fig. 2. The pairing
symmetry of the superconductor is
nodal $p$-wave.
}
  \label{nodalp.fig}
\end{figure}

\begin{figure}
  \centerline{\psfig{figure=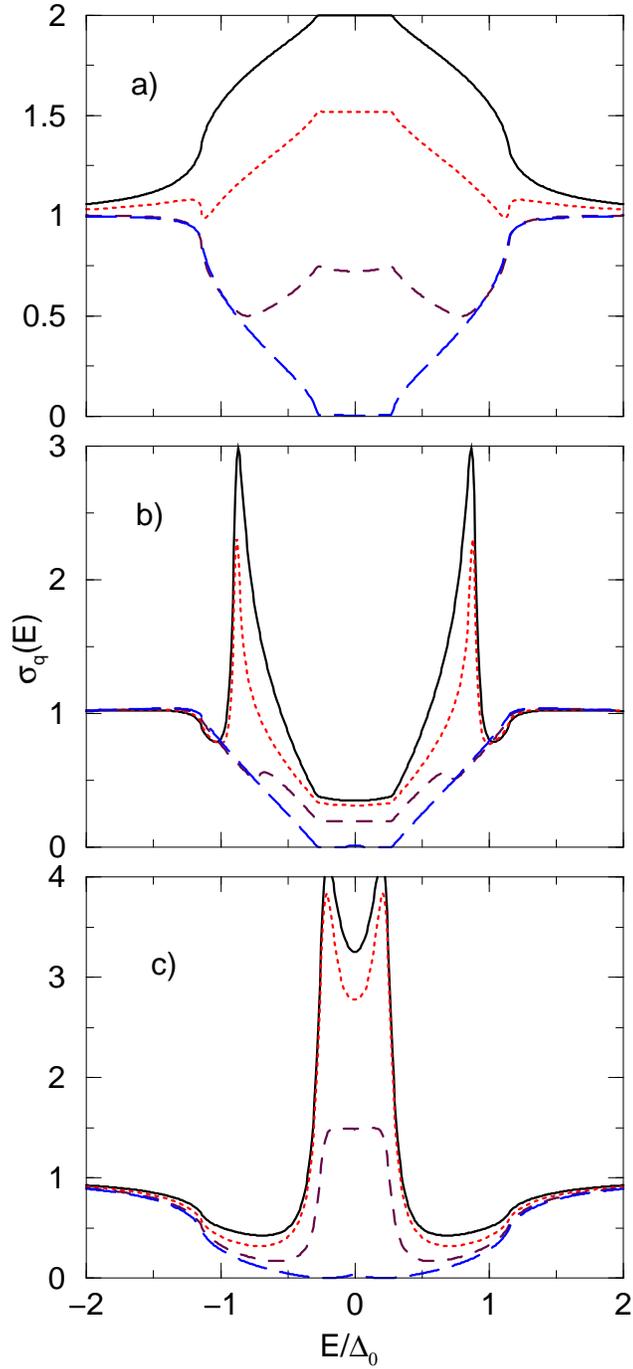,width=8.5cm,angle=0}}
  \caption{
The same as in Fig. 2. The pairing
symmetry of the superconductor is
$p$-wave proposed by M.N.
}
  \label{MN.fig}
\end{figure}

\begin{figure}
  \centerline{\psfig{figure=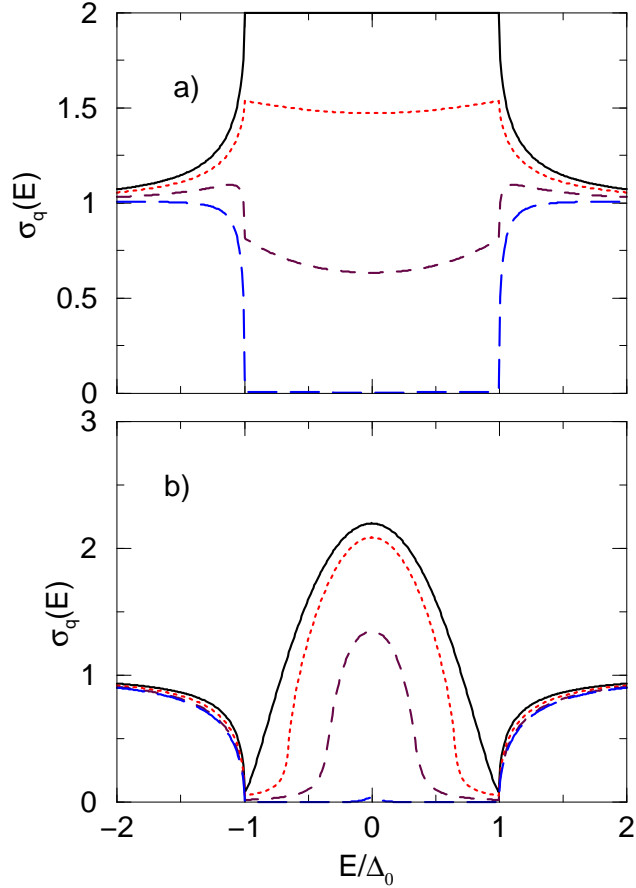,width=8.5cm,angle=0}}
  \caption{
Normalized tunneling conductance $\sigma_q(E)$ as a function of $E/\Delta_0$
for $x=0$ (solid line), $x=0.4$ (dotted line), $x=0.8$ (dashed line), 
and $x=0.999$ (long dashed line), 
for different orientations (a) $z_0=0$, $\beta=0$, (b)$z_0=2.5$,  $\beta=0$.
The pairing 
symmetry of the superconductor is 
isotropic $p$-wave.
}
  \label{iso.fig}
\end{figure}

In Figs. \ref{cos.fig} - \ref{iso.fig}
we plot the tunneling conductance $\sigma_q(E)$  
for different values of the exchange interaction $x=U/E_F$
(a) $z_0=0$, $\beta=0$, 
(b) $z_0=2.5$, $\beta=0$, 
(c) $z_0=2.5$, $\beta=\pi/4$. The pairing 
symmetry of the superconductor is 
$B_{1g}\times E_u$ in Fig. \ref{cos.fig}, 
$B_{2g}\times E_u$ in Fig. \ref{sin.fig}, 
nodal $p$-wave, in Fig. \ref{nodalp.fig}, 
$p$-wave, proposed by K. Miyake, and O. Narikiyo 
\cite{miyake} in Fig. \ref{MN.fig},
and isotropic $p$-wave in Fig. \ref{iso.fig}. 
When the ferromagnet is a normal metal i.e. $x=0$ 
the results of \cite{stefan} are reproduced.
For $z_0=0$, the subgap conductance 
is suppressed, with the increase of $x$, as in the 
case of a $d_{x^2-y^2}$-wave superconductor. \cite{kashiwaya}  

In the case of normal metal / insulator / triplet superconductor
junction 
the peaks inside 
the gap, are connected to bound states, 
which are formed due to the sign change that the 
transmitted quasiparticles feel,
for fixed $\beta$ at discrete values of $\theta$.
The conductance peaks occurs at these energies where 
an increased number of bound states is formed. 
\cite{stefan}.
For unitary pairing states the spins of the 
incident electron and the Andreev reflected hole 
are opposite and since 
the spin up and spin down quasiparticles 
have equal wave vectors, 
no spin effects are involved in the Andreev reflection. 
This is not true, when the 
normal metal is replaced by a ferromagnet. In that case 
the spin-up and spin-down wave vectors are not equal 
and the spin affects the Andreev reflection. 
The Andreev reflected hole decays in the ferromagnet 
and the interference with the reflected electron is weak. 
Moreover the transmitted quasiparticles
experience weakly the sign change of the pair potential 
which is the reason for the formation of the conductance peak. 
Due to this the conductance peaks are suppressed. 
This is seen in Fig. \ref{cos.fig} b(c), $z_0=2.5$ 
and $\beta=0(\pi/4)$, for the $B_{1g} \times E_u$ case.
Quantitatively the suppression of the conductance peaks 
in the ferromagnet / insulator / triplet superconductor 
junction can be seen if we calculate the magnitude 
of the 
Andreev reflected amplitude as a function of the exchange 
field when a bound state is formed. Then the amplitude 
would decay to zero 
with the increase of the exchange field. This calculation 
has been done in the case of a 
ferromagnet / insulator / time reversal symmetry broken superconductor 
junction where simple arguments have been derived to connect 
the suppression of the Andreev reflection as $x$ increases 
with the reduction of the conductance peaks.
\cite{stefan2}

Also in the metallic limit ($x=0$) for all the pairing states 
we see the presence of a large residual 
density of states within the energy gap as a signature 
of unconventional pairing symmetry with higher than two 
angular momentum. This is modified by the presence of the 
ferromagnet, where the increase of the exchange field 
suppresses the density of states within the gap.
Also in the metallic limit the conductance increases
linearly with $E$, which  
is consistent with the presence on line nodes in the 
pairing potential. This linear form of the spectra remains 
unchanged when increasing $x$.
Generally the evolution of the conductance spectra 
with the increase of $x$, for the three pairing symmetries 
with line nodes depends strongly on the position of the 
nodes in the pairing potential.
In the $d_{x^2-y^2}$-wave case 
the peak at $E=0$ and $\beta=\pi/4$, 
due to the sign change of the pairing potential for the 
transmitted quasiparticles, is largely reduced with the increase 
of $x$ \cite{kashiwaya}. In the case of $B_{1g}\times E_u$-wave 
state for $x=0$, the pairing potential 
is more complicated and the sign change occurs at discreet values 
of $\theta$, for fixed $\beta$. 
As a result bound states are formed within 
the gap, and also the position of the conductance peaks depends 
on the orientation angle $\beta$, as seen in Fig. \ref{cos.fig}.
Also the spectra for angle $\beta$ in the $B_{1g}\times E_u$
case is identical to the spectra of $B_{2g}\times E_u$ in Fig. \ref{sin.fig}
for 
angle $\pi /4-\beta$, since the nodes for the two symmetries 
differ by $\pi /4$.
The evolution with $x$ of the conductance spectra for $z_0=0$, is 
different in the $B_{1g}\times E_u$, $B_{2g}\times E_u$ cases. 
It develops a dip at $E=0$ in the $B_{1g}\times E_u$, while it 
has a peak in the $B_{2g}\times E_u$ case, at $E=0$.
The nodal $p$-wave case in Fig. \ref{nodalp.fig} 
has the same nodal structure as the 
$B_{2g}\times E_u$ case and we see that the spectra for these two 
candidates are similar. 

For nodeless pairing states a subdap or a full gap opens in the tunneling 
spectra.
This is seen in Fig. \ref{MN.fig}, 
for the pairing state proposed by K. Miyake, and O. Narikiyo (MN)
\cite{miyake}. The spectra is  similar to the 
nodal $p$-wave case, except that in the (MN) case a subgap 
opens in the tunneling spectra for certain junction orientation. 
In this region for $z_0=0$ the tunneling conductance is 
equal to $2$, when $x=0$ and has a constant value for $x>0$.
The tiny subgap is an indication of nodeless pairing state.
For the isotropic $p$-wave case the tunneling spectra changes 
with $z_0$, as can be seen in Fig. \ref{iso.fig}, but not 
with the boundary orientation $\beta$. The spectra is nodeless, and
for $z_0=0$, the conductance is $\sigma_q(E)=2$, within the energy gap, for 
$x=0$.
Similar results have been obtained in Ref. 
\cite{yoshida}, where the tunneling conductance of 
a ferromagnet / triplet superconductor interface is calculated, for 
both unitary and non-unitary pairing state, having $E_u$ symmetry. 
Generally in all pairing states the reduction of the subgap 
conductance with the exchange field is symmetric since the 
density of states modulation within the subgap 
is not induced by spin dependent effects.
\section{magnetic field effects}

\begin{figure}
  \centerline{\psfig{figure=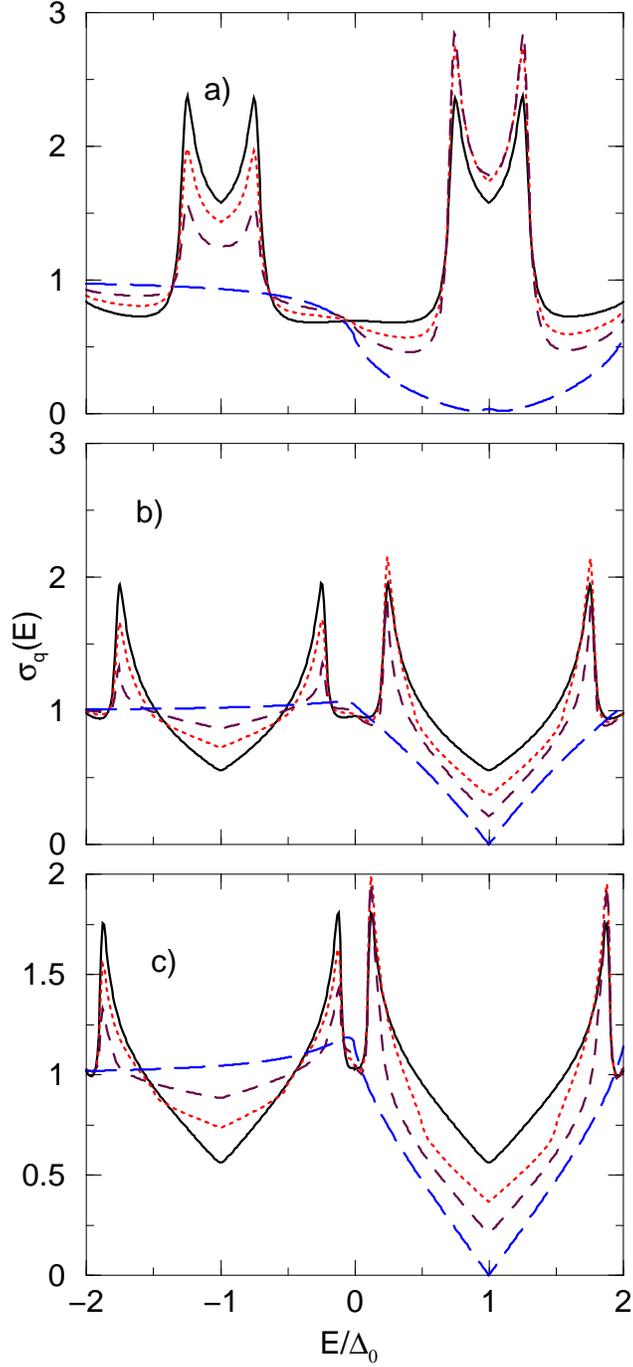,width=8.5cm,angle=0}}
  \caption{
Normalized tunneling conductance $\sigma_q(E)$ as a function of $E/\Delta_0$
for $x=0$ (solid line), $x=0.2$ (dotted line), $x=0.4$ (dashed line), 
and $x=0.999$ (long dashed line), for $z_0=2.5$, $\beta=0.0$,
for different nodal pairing states
(a) $B_{1g}\times E_u$, 
(b) $B_{2g}\times E_u$, 
(c) nodal $p$-wave. 
}
  \label{nodal.fig}
\end{figure}

\begin{figure}
  \centerline{\psfig{figure=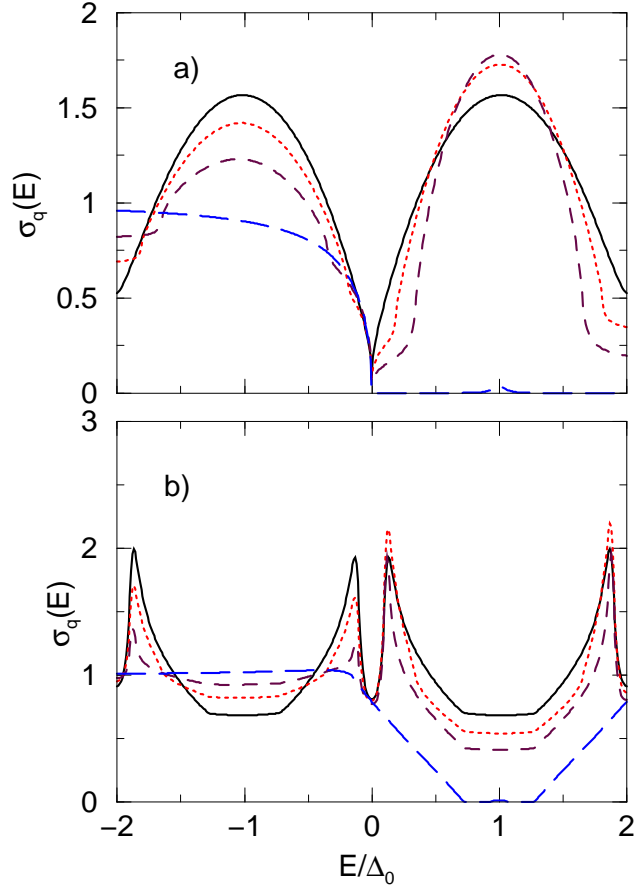,width=8.5cm,angle=0}}
  \caption{
Normalized tunneling conductance $\sigma_q(E)$ as a function of $E/\Delta_0$
for $x=0$ (solid line), $x=0.2$ (dotted line), $x=0.4$ (dashed line), 
and $x=0.999$ (long dashed line), for $z_0=2.5$, $\beta=0.0$,
for different nodeless pairing states
(a) isotropic $p$-wave, 
(b) nodeless $p$-wave proposed by M.N.
}
  \label{nodeless.fig}
\end{figure}

In this section we describe the effect of the external magnetic
field $H$ in the spectra for different values of the exchange field $x$.
We will see that since the effect of the magnetic field depends on the 
spin, the evolution of the tunneling spectra with $x$ is asymmetric.
The tunneling conductance is given by 
\begin{equation}
\sigma_q(E)= \sigma_{q_{\uparrow}}(E-\mu_B H)+
\sigma_{q_{\downarrow}}(E+\mu_B H).
\end{equation}
In Fig. \ref{nodal.fig}a,b,c the tunneling conductance $\sigma_q(E)$ 
is plotted for fixed magnetic field $\mu_B H/\Delta_0=1$, and 
barrier strength $z_0=2.5$, for 
different values of the exchange interaction $x$.
The pairing
symmetry of the superconductor is
$B_{1g}\times E_u$,
$B_{2g}\times E_u$,
nodal $p$-wave, respectively. 
The same information is plotted in Fig. \ref{nodeless.fig} a,b, for the 
nodeless pairing states, isotropic $p$-wave, and 
$p$-wave, proposed by K. Miyake, and O. Narikiyo,\cite{miyake} 
respectively. The orientation of the superconductor is chosen 
as $\beta=0$. 
In the absence of the exchange interaction $(x=0)$ the 
magnetic field splits symmetrically the tunneling spectrum. The 
amplitude of the spiting depends linearly on the magnetic field $H$.
The main effect of the polarization is the imbalance 
in the peak heights for $E$ positive and negative. 
The ratio of the peaks for positive and negative energy 
is proportional to the exchange field of the material.
For $E<0$ the pattern is suppressed linearly with the 
increase of the $x$, while for $E>0$ the tunneling 
conductance spectra initially increases with $x$ and 
then decreases.  
Also the conclusions of the previous section for the 
nodal (nodeless) form of the tunneling spectra 
are still valid in the presence of a magnetic field.

\section{Conclusions} 
We calculated the tunneling conductance in ferromagnet / insulator / 
triplet superconductor junctions, using the BTK formalism. We assumed 
pairing potentials with nodes such as 
the nodal $p$-wave and two $2D$
$f$-wave states with line nodes, breaking the time reversal 
symmetry.
Also we examined two nodeless pairing states, the $p$-wave 
proposed by K. Miyake, and O. Narikiyo
\cite{miyake}, and the isotropic $p$-wave. 
The linear variation of the conductance with $E$ 
is an indication of line nodes and is not influenced 
much when the exchange interaction increases. 
On the other hand the large residual density of states 
within the gap is reduced with the increase of $x$, 
and the peaks due to the formation of bound states 
are removed due to the suppression of the Andreev reflection. 
The evolution of the spectra with $x$ depends on the 
position of the nodes and is different in the three pairing 
states with line nodes. 
In the case of nodeless pairing states the tunneling conductance 
develops a subgap or a full gap 
where $\sigma_q(E)$ has a constant value within.
The exchange interaction suppresses the conductance within 
the gap, and can be considered as a measure of the 
polarization of the material.
The magnetic field splits linearly the tunneling spectra and 
in the half metallic ferromagnetic limit $x=1$, eliminates 
the negative branch of the spectrum.
These features can be used to distinguish between
the candidate pairing symmetry states of Sr$_2$RuO$_4$.

\end{document}